\documentstyle[twocolumn,prl,aps]{revtex}

  \newcommand{\beq}{\begin{equation}}
  \newcommand{\eeq}{\end{equation}}
  \newcommand{\beql}[1]{\begin{equation}\label{eq:#1}}
  \newcommand{\beqa}{\begin{eqnarray}}
  \newcommand{\eeqa}{\end{eqnarray}}
  \newcommand{\beqas}{\begin{eqnarray*}}
  \newcommand{\eeqas}{\end{eqnarray*}}                       
  \newcommand{\bA}{{\bf A}}
  \newcommand{\bC}{{\bf C}}
  \newcommand{\bS}{{\bf S}}
  \newcommand{\bT}{{\bf T}}
  \newcommand{\cE}{{\cal E}}
  \newcommand{\al}{\alpha}

  \newcommand{\da}{\dagger}
  \newcommand{\de}{\delta}

  \newcommand{\nn}{\nonumber}

  \newcommand{\ps}{\psi} 
  \newcommand{\rh}{\rho}
  \newcommand{\si}{\sigma}

  \newcommand{\De}{\Delta}                                          
  
  \newcommand{\Eq}[1]{Eq.~(\ref{eq:#1})}

  \newcommand{\Tr}{\mbox{\rm Tr}}

  \newcommand{\eq}[1]{(\ref{eq:#1})}

\newcommand{\bra}[1]{\langle#1|}
\newcommand{\ket}[1]{|#1\rangle}

\newcommand{\bracket}[1]{\langle#1\rangle}
\newcommand{\NOT}{{\rm NOT} }
\newcommand{\SWAP}{{\rm SWAP} }
\newcommand{\CN}{{\rm CN} }

\begin{document}
\draft
\title{Conservative Quantum Computing}
\author{Masanao Ozawa}
\address{Graduate School of Information Sciences,
T\^{o}hoku University, 
Aoba-ku, Sendai 980-8579, Japan\\
Center for Photonic Communication and Computing,
Department of Electrical and Computer Engineering, Northwestern
University, Evanston, IL 60208}
\maketitle

\begin{abstract}
Conservation laws limit the accuracy of physical implementations of 
elementary quantum logic gates.  If the computational basis is represented
by a component of spin and physical implementations obey the angular
momentum conservation law, any physically realizable unitary operators
with size less than $n$ qubits cannot implement the controlled-\NOT gate
within the error probability $1/(4n^{2})$, where the size is defined as the
total number of the computational qubits and the ancilla qubits.
An analogous limit for bosonic ancillae is also obtained to show that the 
lower bound of the error probability is inversely proportional 
to the average number of photons. 
Any set of universal gates inevitably obeys a related
limitation with error probability $O(1/n^{2})$. To circumvent the
above or related limitations yielded by conservation laws, it is
recommended that the computational basis should be chosen as the
one commuting with the additively conserved quantities.
\end{abstract}

\narrowtext

Since the discovery of Shor's algorithm \cite{Sho97}, 
physical realization of quantum computers is one of
the major topics in physics. One of the formidable obstacles to
the realization of quantum computers is the decoherence induced by 
the environment. 
The theory of quantum error correction and the theory of fault-tolerant
quantum computing have been developed to overcome this
difficulty\cite{NC00,note1}. One of the main achievements of this field is
the threshold theorem: Provided the noise in individual quantum gates is
below a  certain threshold it is possible to efficiently perform an arbitrarily
large quantum computing.
However, the threshold 
is rather
demanding and the problem turns to whether there is any 
fundamental limit for implementing quantum gates.
Recently, Lloyd \cite{Llo00} and Ng \cite{Ng01} have discussed 
how fundamental constants provide limits on speed and memory
of quantum computers.  Here, I will propose another approach based
on conservation laws.

If we consider the ultimate performance of computing allowed
by the laws of physics, elementary quantum gates should be isolated and
small, so that the corresponding unitary operators should
satisfy fundamental symmetries, or conservation laws.  From
this point of view, it is likely that the degree of conflict with a conservation
law
depends on the nature of  its logic to be performed and that 
the  imperfection can be reduced by increasing the
size of implementation.  However, no serious investigation has ever taken
place.
In this letter we model qubits as spin 1/2
objects and investigate the quantum limit induced by the angular momentum
conservation law.   We show that although the \SWAP gate has no conflict
with the conservation law, the controlled-\NOT gate, which is one of the 
universal quantum logic gates, 
cannot be
implemented by any 2-qubit rotationally invariant unitary operation within
error probability $1/16$.  Thus, to obtain more accuracy, we need to blow up
the unitary operation to an ancilla system.  Then, the size of an
implementation of the quantum gate is defined as the total number of qubits
in the computational basis and the ancilla.  It is shown that any physically
realizable unitary operator with size less than
$n$ qubits cannot implement the controlled-\NOT gate within the
error probability $1/(4n^{2})$.  
An analogous limit for bosonic ancillae will be also obtained by defining
the size of the ancilla as 2 times the square root of the average number of photons,
and thus the lower bound is inversely proportional to the average number of photons.
 It is also shown that in any set of universal gates, for any size
limit $s$ there is at least one gate which cannot be implemented within the
error probability $1/(ks^{2})$ for some constant $k$.  Thus, we cannot 
circumvent this limitation by a clever choice of the set of universal
gates.

Let $U_{\CN}$ be a controlled-\NOT gate on a 2-qubit system 
$\bC+\bT$.  
Let $X_{i}$, $Y_{i}$, and $Z_{i}$ be the Pauli operators of qubit $\bC$
for $i=1$ or qubit $\bT$ for $i=2$ defined by
$X_{i}=\ket{1}\bra{0}+\ket{0}\bra{1}$,
$Y_{i}=i\ket{1}\bra{0}-i\ket{0}\bra{1}$, and 
$Z_{i}=\ket{0}\bra{0}-\ket{1}\bra{1}$
with the computational basis
$\{\ket{0},\ket{1}\}$.
On the computational basis, $U_{\CN}$ acts as
$U_{\CN}\ket{a,b}=\ket{a,b\oplus a}$ for $a, b=0,1$,
where $\oplus$ denotes the addition modulo 2.
Thus, in particular, we have
\beql{measurement}
U_{\CN}\ket{a,0}=\ket{a,a}
\eeq
for $a=0,1$.
The above relation shows that the unitary operator $U_{\CN}$
serves as an interaction between the ``object'' $\bC$ and the
``probe'' $\bT$ for a measurement of $Z_{1}$ satisfying the 
projection postulate.
Thus, by the Wigner-Araki-Yanase theorem \cite{Wig52,AY60}, if there
are additive conserved quantities not commuting with $Z_{1}$,
the unitary operator
$U_{\CN}$ cannot be implemented.  
To be precise,  let $L_{1}$ and 
$L_{2}$ be a pair of observables of $\bC$ and $\bT$, 
respectively, such that 
\beql{commuting}
[Z_{1},L_{1}]\not=0.
\eeq
Then, the controlled-\NOT gate $U_{\CN}$  cannot satisfy
the conservation law\cite{footnote2}
\beql{conserv}
[U_{\CN},L_{1}+L_{2}]=0.
\eeq
A simple proof runs as follows.
Assume \Eq{conserv}.
If $a\not=b$, we have 
\beqas
\bracket{a|L_{1}|b}
&=&
\bracket{a,0|L_{1}+L_{2}|b,0}\\
&=&
\bracket{a,0|U^{\da}_{\CN}(L_{1}+L_{2})U_{\CN}|b,0}\\
&=&
  \bracket{a,a|L_{1}+L_{2}|b,b}=0.
\eeqas
Thus, $L_{1}$ is diagonal in the computational basis of $\bC$.
Therefore, {\em if $L_{1}$ does not commute with $Z_{1}$, then
$U_{\CN}$ cannot satisfy the conservation law \eq{conserv}.}
In particular, $U_{\CN}$ cannot be implemented in the presence of 
the angular momentum conservation law.

The above impossibility of implementation depends on the logic.
Despite the limitation on the controlled-\NOT gate, the \SWAP gate
$U_{\SWAP}$, defined by 
$
U_{\SWAP}\ket{a,b}=\ket{b,a}
$
for $a,b=0,1$, can be implemented precisely under the angular 
momentum conservation law.  In fact, the \SWAP
gate can be precisely implemented as \cite{OP88}
\beq
U_{\SWAP}
=\exp{\frac{-i\pi}{4}(-1+X_{1}X_{2}+Y_{1}Y_{2}+Z_{1}Z_{2})}.
\eeq

In order to construct a physical implementation of $U_{\CN}$,
the above consideration suggests the need for blowing up the unitary
operation to a larger system including additional qubits.
Let $\al=(U,\ket{\xi})$ be a physical implementation of $U_{\CN}$ defined
by a unitary operator $U$ on the system $\bC+\bT+\bA$, where $\bA$ is
a quantum system called the {\em ancilla}, and a state vector
$\ket{\xi}$ of the ancilla, in which the ancilla is prepared
at the time at which $U$ is turned on.  
The implementation $\al=(U,\ket{\xi})$ defines a 
trace-preserving quantum operation $\cE_{\al}$ by
\beq
\cE_{\al}(\rh)=\Tr_{\bA}[U(\rh\otimes\ket{\xi}\bra{\xi})U^{\da}]
\eeq
for any density operator $\rh$ of the system $\bC+\bT$, 
where $\Tr_{\bA}$ stands for the partial trace over the system $\bA$.
On the other hand, the gate $U_{\CN}$ defines the trace-preserving
quantum operation ${\rm ad}U_{\CN}$ by 
\beq
{\rm ad}U_{\CN}(\rh)=U_{\CN}\rh U_{\CN}^{\da}
\eeq
for any density operator $\rh$ of the system $\bC+\bT$.

How successful the implementation $(U,\ket{\xi})$
has been is most appropriately measured by the 
{\em completely bounded (CB) distance}\cite{note-CB-distance}
between two operations $\cE_{\al}$ and ${\rm ad}U_{\CN}$ defined by
\beql{CB-distance}
D_{CB}(\cE_{\al},U_{\CN})=\sup_{n,\rh}
D(\cE_{\al}\otimes {\rm id}_{n}(\rh),{\rm ad}U_{\CN}\otimes {\rm id}_{n}(\rh)),
\eeq
where $n$ runs over positive integers, ${\rm id}_{n}$ is the identity operation
on an $n$-level system $\bS_{n}$,  $\rh$ runs over density operators
of the system $\bC+\bT+\bS_{n}$, and $D(\si_{1},\si_{2})$ stands 
for the trace distance\cite[p.~403]{NC00} of two states $\si_{1}$ and
$\si_{2}$. Since the trace distance of the above two states can be interpreted
as an achievable upper bound on the so-called total variation distance of two
probability distributions arising from measurements performed on the two
output states of the corresponding gates \cite[p.~405]{NC00}, we interpret
$D_{CB}(\cE_{\al},U_{\CN})$ as the worst error probability of
operation $\cE_{\al}$ in simulating the gate $U_{\CN}$ on any 
input state of any circuit including those two gates.  We shall call 
$D_{CB}(\cE_{\al},U_{\CN})$ the {\em gate error probability} of the
implementation $\al$ of the gate $U_{\CN}$. 

Another measure, which is more tractable in computations,
is the {\em gate fidelity}\cite[p.~418]{NC00} defined by
\beq
F(\cE_{\al},U_{\CN})
=\min_{\ket{\ps}}F(\ps)
\eeq
where $\ket{\ps}$ varies over all state vectors of $\bC+\bT$, and 
$F(\ps)$
is the fidelity of two states 
$U_{\CN}\ket{\ps}$ and $\cE_{\al}(\ket{\ps}\bra{\ps})$ given by
\beql{fidelity}
F(\ps)=
\bracket{\ps|U_{\CN}^{\da}\cE_{\al}(\ket{\ps}\bra{\ps})U_{\CN}|\ps}^{1/2}.
\eeq
By the relation\cite[p.~416]{NC00} 
\beql{lower-bound}
1-F(\cE_{\al},U_{\CN} )^{2}\le D_{CB}(\cE_{\al},U_{\CN}),
\eeq
any lower bound of $1-F(\cE_{\al},U_{\CN} )^{2}$ gives a
lower bound of the gate error probability.
The operator $U$ and the operation $\cE_{\al}$ is generally
described by the following actions on computational basis states
\beqa
U\ket{a,b}\ket{\xi}&=&\sum_{c,d=0}^{1}\ket{c,d}\ket{E^{ab}_{cd}}\\
\cE_{\al}(\ket{a,b}\bra{a,b})
&=&\sum_{i,j,k,l=0}^{1}\ket{i,j}
\bracket{E^{a,b}_{k,l}|E^{a,b}_{i,j}}\bra{k,l}
\eeqa
for $a,b=0,1$, where $\ket{E^{ab}_{cd}}$ is not
necessarily normalized.
It follows that the fidelity is given by 
\beq\label{eq:fidelity-values}
F(a,b)=\|\ket{E^{a,b}_{a,b\oplus a}}\|.
\eeq

Now, we assume that there are additive conserved quantities 
$L_{1}$,  $L_{2}$, and $L_{3}$ of systems $\bC$, $\bT$, 
and $\bA$, respectively, so that the unitary operator $U$ should satisfy
the conservation law
\beql{conservation}
[U,L_{1}+L_{2}+L_{3}]=0.
\eeq
Since computational qubits, $\bC$ and $\bT$, should have the same 
physical structure, we naturally assume
$\|L_{1}\|=\|L_{2}\|$ for their operator norms.

Our problem is to find a good lower bound of the gate error probability
\eq{CB-distance} under the conservation law \eq{conservation}.
In order to derive the lower bound from uncertainty relations,  
we introduce the deviation operators $D_{ij}$ of the system
$\bC+\bT+\bA$ for $i,j=1,2$ defined by 
\beq\label{eq:noise-operators}
D_{ij}={Z}'_{i}-{Z}_{j},
\eeq 
where we write ${A}'=U^{\da}AU$
for any operator $A$.
The root-mean-square deviation $\de_{ij}(\ps)$ on arbitrary input
state $\ket{\ps}$ of $\bC$ is defined as the root-mean-square of the
deviation operator
$D_{ij}$ in state $\ket{\ps,0,\xi}=\ket{\ps}\ket{0}\ket{\xi}$, i.e., 
\beq
\de_{ij}(\ps)=\bracket{D_{ij}^{2}}^{1/2},
\eeq
where $\bracket{\cdots}$ abbreviates 
$\bracket{\ps,0,\xi|\cdots|\ps,0,\xi}$.
For any observable $A$, 
we shall denote by $\De A$  the standard deviation of $A$ defined by
$\De A=\bracket{(A-\bracket{A})^{2}}^{1/2}$.  
Then, we easily see
\beql{E>D}
\De D_{ij}\le\de_{ij}(\ps)
\eeq
for $i,j=1,2$.
In the case where $U=U_{\CN}$, we have $D_{11}=0$,
$D_{12}=Z_{1}-Z_{2}$,
$D_{21}=Z_{1}(Z_{2}-I)$, and 
$D_{22}=(Z_{1}-I)Z_{2}$, so that
$\de_{11}(\ps)=\de_{21}(\ps)=0$ for any
state $\ket{\ps}$ of $\bC$.  Thus, the relation $\de_{11}(\ps)^{2}+
\de_{21}(\ps)^{2}>0$ implies $U\not=U_{\CN}$. Hence, the quantity
$\de_{11}(\ps)^{2}+\de_{21}(\ps)^{2}$  measures a degree of
imperfection. 

Now, we shall evaluate $\de_{11}(\ps)$ and $\de_{21}(\ps)$
for a general implementation $\al=(U,\ket{\xi})$ under the conservation
law \eq{conservation}.
From the conservation law \eq{conservation} and the relations
$[Z_{1},L_{2}]=[Z_{1},L_{3}]=0$, we have
\beq
[{Z}_{1},{L}_{1}]
=[{Z}_{1},{L}_{1}']+[{Z}_{1},{L}_{2}']
+[{Z}_{1},{L}_{3}'].
\eeq
From the definition of deviation operators, \Eq{noise-operators}, we have
\begin{mathletters}\beqa
&[{Z}_{1},{L}_{1}']=[{L}_{1}',D_{21}]\quad\mbox{and}\quad
[{Z}_{1},{L}_{2}']=[{L}_{2}',D_{11}],&\\{}
&[{Z}_{1},{L}_{3}']=[{L}_{3}',D_{11}]
                                               =[{L}_{3}',D_{21}].&
\eeqa\end{mathletters}
Thus, we have the following noise commutation relations
\beqa\label{eq:noise-commutation1}
[{Z}_{1},{L}_{1}]
&=&
[{L}_{1}',D_{21}]+[{L}_{2}',D_{11}]+[{L}_{3}',D_{11}],\\
\label{eq:noise-commutation2}
[{Z}_{1},{L}_{1}]
&=&
[{L}_{1}',D_{21}]+[{L}_{2}',D_{11}]+[{L}_{3}',D_{21}].
\eeqa
Taking the modulus of the expectations of the both sides of  
\Eq{noise-commutation1}
and applying the triangular inequality, we have
\beqa
\lefteqn{|\bracket{[Z_{1},L_{1}]}|}\quad\nn\\
&\le&
|\bracket{[{L}_{1}',D_{21}]}|
+|\bracket{[{L}_{2}',D_{11}]}|
+|\bracket{[{L}_{3}',D_{21}]}|.
\eeqa
By the uncertainty relation \cite{Rob29} and \Eq{E>D}, we have
\beq
|\bracket{[L_{k}',D_{ij}]}|\le 2\De D_{ij}\De L'_{k}
\le 2\de_{ij}(\ps)\De L'_{k}.
\eeq
Thus, we obtain the following consequence of the first noise commutation
relation, \Eq{noise-commutation1},
\beqa
|\bracket{[Z_{1},L_{1}]}|
&\le&
  2\de_{21}(\ps)\De{L}_{1}'
+2\de_{11}(\ps)\De{L}_{2}'\nn\\
& &+2\de_{11}(\ps)\De{L}_{3}'.
\eeqa
Similarly, from the second noise commutation relation,
\Eq{noise-commutation2}, we obtain the following relation
\beqa
|\bracket{[Z_{1},L_{1}]}|
&\le&
  2\de_{21}(\ps)\De{L}_{1}'
+2\de_{11}(\ps)\De{L}_{2}'\nn\\
& &+2\de_{21}(\ps)\De{L}_{3}'.
\eeqa
Summing up both inequalities and using the relations
$\De{L}_{1}',\De{L}_{2}'\le
\|L_{1}\|=\|L_{2}\|$, we have
$$
|\bracket{[Z_{1},L_{1}]}|
\le
(\de_{11}(\ps)+\de_{21}(\ps))(2\|L_{1}\|+\De L'_{3}).
$$
By the inequality
$(x+y)^{2}/2\le x^{2}+y^{2}$, we have
the lower bound of the imperfection
\beql{squared-noise-bound}
\frac{|\bracket{[Z_{1},L_{1}]}|^{2}}
{2(2\|L_{1}\|+\De L'_{3})^{2}}\le
\de_{11}(\ps)^{2}+\de_{21}(\ps)^{2}.
\eeq

Let us consider the computational basis defined by the
spin component of the $z$ direction and 
the angular momentum conservation law for the $x$ direction.
Thus, we assume $L_{i}=X_{i}$ for
$i=1,2$, so that 
\beql{norm}
\|L_{1}\|=\|L_{2}\|=1,
\eeq
and that $L_{3}$ is considered as the $x$-component of 
the total angular momentum divided by $\hbar/2$
of the ancilla system $\bA$.
In order to maximize the bound in 
\Eq{squared-noise-bound}, suppose that the input state
$\ket{\ps}$ is the spin state of the $y$ direction, i.e.,
$
\ket{\ps}=\frac{1}{\sqrt{2}}(\ket{0}+\ket{1}).
$
Then, by straightforward calculations we have
\beqa
\de_{11}(\ps)^{2}
&=&2\|\ket{E^{10}_{00}}\|^{2}
+2\|\ket{E^{10}_{01}}\|^{2}
+2\|\ket{E^{00}_{10}}\|^{2}\nn\\
& &+2\|\ket{E^{00}_{11}}\|^{2},\\
\de_{21}(\ps)^{2}&=&2\|\ket{E^{10}_{00}}\|^{2}
+2\|\ket{E^{00}_{01}}\|^{2}
+2\|\ket{E^{10}_{10}}\|^{2}\nn\\
& &+2\|\ket{E^{00}_{11}}\|^{2}.
\eeqa
Since $\sum_{c,d=0}^{1}\|\ket{E^{ab}_{cd}}\|^{2}=1$ for
$a,b=0,1$, from
\Eq{fidelity-values} we have
\beqa\label{eq:fidelity-bound}
\lefteqn{\de_{11}(\ps)^{2}+\de_{21}(\ps)^{2}}\nn\qquad\\
&\le&4[1-F(00)^{2}]+4[1-F(10)^{2}]\nn\\
&\le&8[1-F(\cE_{\al},U_{\CN})^{2}].
\eeqa
Since $[Z_{1},L_{1}]=[Z_{1},X_{1}]=2iY_{1}$, we have
\beql{commutativity}
|\bracket{[Z_{1},L_{1}]}|=2.
\eeq
Thus, from Eqs.~\eq{squared-noise-bound}, \eq{norm},
\eq{fidelity-bound}, and \eq{commutativity}, we have the following
fundamental lower bound of the gate error probability 
\beq
\frac{1}
{4(2+\De L'_{3})^{2}}
\le
1-F(\cE_{\al},U_{\CN})^{2}
\le D_{CB}(\cE_{\al},U_{\CN}).
\eeq
In the following, we shall interpret the above relation in terms
of the notion of the size of implementations for
fermionic and bosonic ancillae separately.

We now assume that the ancilla $\bA$ comprises qubits.
Then, the size $s(\al)$ of the implementation $\al$ is defined to be
the total number $n$ of the qubits included in $\bC+\bT+\bA$.
Then, we have
\beq
\De L'_{3}\le \|L_{3}\|=n-2.
\eeq
Thus, we have the following lower bound of the gate
error probability
\beql{bound}
\frac{1}{4s(\al)^{2}}\le 1-F(\cE_{\al},U_{\CN})^{2}
\le D_{CB}(\cE_{\al},U_{\CN}),
\eeq
with $s(\al)=n$.
Therefore, it has been proven that {\em if the computational basis is 
represented by the $z$-component of spin, any implementation
with size $n$ which preserves the $x$-component of angular
momentum cannot simulate the controlled-\NOT gate within
the error probability $1/(4n^{2})$}.  In particular, any implementation
on $\bC+\bT$ cannot simulate $U_{\CN}$ within the error probability
$1/16$.

In current proposals 
\cite{NC00,note1}, the external electromagnetic field prepared by the laser 
beam is considered to be a feasible candidate for the ancilla 
$\bA$ to be coupled with the computational qubits $\bC+\bT$ via the
dipole interaction.  In this case, an analogous limit for bosonic ancillae
is obtained by defining the size of the ancilla as 2
times the square root of the average number of photons, and
thus the lower boud is inversely proportional to the average
number of phtons.  In fact, the ancilla state 
$\ket{\xi}$ is considered to be a coherent state, for which we have 
$(\De N)^{2}=\bracket{\xi|N|\xi}=\bracket{N}$, where $N$ is the number
operator.  We assume that the beam propagates to the
$x$-direction with RHC polarization.  Then, we
have
$L_{3}=2N$,  and hence
\beq
\De L'_{3}=2\De N'=2\bracket{N'}^{1/2}\le 2(\bracket{N}+2)^{1/2}.
\eeq
Thus, \Eq{bound} holds with defining the size of implementation
$\al$ by
$s(\al)=2\bracket{N}^{1/2}$ appropriately for the strong field, and hence \Eq{bound}
turns to be the relation 
\beql{bound2}
\frac{1}{16\bracket{N}}\le 1-F(\cE_{\al},U_{\CN})^{2}
\le D_{CB}(\cE_{\al},U_{\CN}).
\eeq

Formula \eq{bound} holds, therefore, appropriately for both fermionic
and bosonic ancillae.
In the
most general case, \Eq{bound} holds with $s(\al)=2+\De
L'_{3}$ dependent on the ancilla state, or with $s(\al)=2+\|L_{3}\|$ 
independent of the ancilla state.

The above limit on implementations of elementary
gates cannot be circumvented by any choices of the set of universal
gates.  In fact, we can generally prove that {\em in any set of universal gates,
for any size limit $s$ there is at least one gate which cannot be
implemented within the error probability
$1/(ks^{2})$ for some constant $k$}.
A proof runs as follows.
Suppose that $U_{\CN}$ can be constructed from $m$ elementary gates.
Let $U_{\CN}=U_{m}\cdots U_{1}$ and $\cE_{\al}=\cE_{m}\cdots
\cE_{1}$, where $\cE_{i}$ is the operation of the best implementation
of gate $U_{i}$ with size $s$.  Then, $s(\al)\le ms$, and
hence from the chain property of CB distance\cite{NC00,AKN98},  
we have
\beq
\frac{1}{4(ms)^{2}}\le
D_{CB}(\cE_{\al},U_{\CN})
\le\sum_{i=1}^{m}D_{CB}(\cE_{i},U_{i}).
\eeq
Thus, one of $U_{i}$ must satisfy
$
1/(4m^{3}s^{2})\le D_{CB}(\cE_{i},U_{i}).
$

By modifying the model of a measurement 
due to Araki and Yanase \cite{AY60}, 
it can be shown that there is a physical 
implementation $\al$ of $U_{\CN}$ 
with any size $n$ satisfying the angular momentum conservation law 
such that $1-F(\cE_{\al},U_{\CN})^{2}=O(1/n)$.
Thus, it is really possible to make the error probability small by 
making the ancilla large.  The detailed construction 
will be discussed elsewhere.

Although it is difficult to envisage what the hardware of the
quantum computer will be like, in order to realize a mobile quantum 
computer a fermionic ancilla appears to be plausible.
The current theory 
demands the ``threshold" error probability $10^{-5}$--$10^{-6}$ 
for each quantum gate \cite[p.~482]{NC00}.  
Thus, a single controlled-\NOT gate would not in reality 
a unitary operation on a 2-qubit system but will be 
a unitary operation on a system with at
least 100 qubits, as long as the computational basis is chosen as a spin
component. The present investigation suggests that the current choice
of  the computational
basis should be modified so that the computational basis commutes
with the conserved quantity.  
Since the additive conserved quantity has degenerate spectrum on the
multiple qubits, we may find such a computational basis comprised 
of orthogonal entangled states over a multiple-qubit system.  
Accordingly,  the theory of fault-tolerant quantum
computing based on single qubit errors should be modified to
incorporate with such choice of the computational basis.

\acknowledgments

The author thanks Julio Gea-Banacloche and 
Horace Yuen for helpful comments.
This work was supported by the R\&D on Quantum 
Commun.\ Tech.\ Program of MPHPT, 
by the CREST project of the JST,
and by the Grant-in-Aid for Scientific Research of the JSPS.


\begin{references}
\bibitem{footnote2}
We denote the operator extended to a larger system by the same symbol as
the original, e.g., $L_{1}$ instead of $L_{1}\otimes 1$.
\end{references}
\end{document}